\documentclass[twocolumn,preprintnumbers,amsmath,amssymb,floatfix]{revtex4}
\usepackage{graphicx}% Include figure files
\usepackage{dcolumn}% Align table columns on decimal point
\usepackage{bm}% bold math
\usepackage{epsf}
\usepackage{amsmath}
\usepackage{amsfonts}
\newcommand{\DDir}{\relax{D\kern-.7em{/}}}

\newcommand{\inv}[1]{\frac{1}{#1}}

\newcommand{\be}{\begin{equation}}
\newcommand{\ee}{\end{equation}}
\newcommand{\bea}{\begin{equation*}}
\newcommand{\eea}{\end{equation*}}

\newcommand{\nin}{\relax{\in\kern-.8em{/}}}

\newcommand{\te}{\theta}

\newcommand{\cm}{\mbox{ cm}}

\newcommand{\erg}{\mbox{ erg}}

\newcommand{\GHz}{\mbox{ GHz}}

\newcommand{\Mpc}{\mbox{ Mpc}}

\newcommand{\mJy}{\mbox{ mJy}}

\newcommand{\sigSIC}{r_{SIC}}
\renewcommand{\apj}{ApJ}

\newcommand{\mnras}{MNRAS}
\begin{document}
\title{Radius and magnetic field from Synchrotron-self-absorbed radio and Inverse Compton X-ray observations of Supernovae}
\author{Boaz Katz$^{1 *}$}
\affiliation{$^1$Institute for Advanced Study, Princeton, NJ 08540, USA}
\begin{abstract}
Simple expressions for the radius and magnetic field of a system emitting Synchrotron-self-absorbed radio and Inverse Compton X-rays are derived from first principles which involve observable quantities only. These expressions are useful for analyzing observations of Supernova blastwaves interacting with dense circumstellar material at early times. 
\end{abstract}
\maketitle

Expressions for the radius and magnetic field of a system emitting Synchrotron-Self-Absorbed (SSA) radio were derived in \cite{Scott77, Chevalier98} and applied to Supernova blastwaves interacting with dense circumstellar material in \citep{Slysh90, Chevalier98} . These expressions (weakly) depend on the ratio of the unknown energy densities of relativistic electrons and the magnetic field. Here we derive expressions that are parameter independent for the abundant cases \citep[e.g.][]{Chevalier06} where Inverse Compton (IC) emission is also observed.
  
Consider a spherical source of radius $R$ observed at a distance $d$, with a power law distribution of electrons $dN/d\gamma\propto \gamma^{-p}$, magnetic field $B$ (isotropically oriented) and target photon energy density $U_t$ with observed flux $F_t=U_tc\te^2$ where $\te=R/d$ is the angular scale of the source on the sky. The SSA flux per logarithmic frequency reaching the observer is given by \citep[e.g.][]{Rybicki79}:
\begin{align}\label{eq:SelfAbsSyn}
f_{\nu,\rm abs}=\pi D(p)\te^2\frac{2\nu^{2}}{3c^2}\left(\frac{4\pi\nu m_e c}{eB}\right)^{1/2}m_ec^2
\end{align}
with $D(p)$ given by
\begin{equation*}
D(p)=\inv{\sqrt{6}(p+1)}\frac{\Gamma\left(\frac{3p-1}{12}\right)\Gamma\left(\frac{3p+19}{12}\right)}{\Gamma\left(\frac{3p+2}{12}\right)\Gamma\left(\frac{3p+22}{12}\right)}\frac{\Gamma\left(\frac{p+5}{4}\right)\Gamma\left(\frac{p+8}{4}\right)}{\Gamma\left(\frac{p+7}{4}\right)\Gamma\left(\frac{p+6}{4}\right)},
\end{equation*}
and equal to 
\begin{equation*}
\begin{array}{cccccccc}
p&=&2  &2.2 &2.4  &2.6  &2.8  &3\\
D(p)&=&0.2010    &0.1773  &0.1581    &0.1423    &0.1291    &0.1179.
\end{array}
\end{equation*}
Equation \eqref{eq:SelfAbsSyn} is equivalent to the Rayleigh Jeans tail of black-body emission with a frequency dependent temperature, $T_{\nu}=D(p)[4\pi\nu m_e c/(eB)]^{1/2}m_ec^2/3$, corresponding to the effective energy of the electrons emitting Synchrotron at the frequency $\nu$. 
The radio emission at high frequencies, where the system is optically thin, is related to the IC X-ray emission from the same electrons by
\begin{equation}\label{eq:SynIC}
\frac{B^2}{8\pi U_{t}}=\frac{\nu f_{\nu,syn}}{\nu f_{\nu,IC}}\equiv \sigSIC.
\end{equation}
For the most widely observed case $p=3$ \citep[e.g.][]{Chevalier06}, where $\nu f_{\nu}=$const, these do not need to be the same electrons.
There are thus two equations, \eqref{eq:SelfAbsSyn} and \eqref{eq:SynIC}, for two variables $B$, and $\te$, which can be solved for in terms of the observable parameters, $\sigSIC$, $F_t$ and $f_{\nu,\rm abs}$. 

Using the following combination of observable quantities,
\begin{equation*}
B_t=(8\pi \sigSIC F_t/c)^{1/2},~~~~f_{B_t}=\pi D(p)\frac{2\nu^{2}}{3c^2}\left(\frac{4\pi\nu m_e c}{eB_t}\right)^{1/2}m_ec^2,
\end{equation*}
we find 
\begin{equation*}
\frac{R}{d}=\te=\left(\frac{f_{\nu,\rm abs}}{f_{B_t}}\right)^{2/5},~~~B=B_t\te^{-1},
\end{equation*}
or
\begin{align*}
R\approx &1.86\times 10^{15}\cm \left(\frac{D(p)}{D(3)}\right)^{-2/5}\frac{d}{10\Mpc}\cr
&\times\left(\frac{\sigSIC F_t}{10^{-12}\erg\cm^{-2}\sec^{-1}}\right)^{1/10}\left(\frac{f_{\nu,\rm abs}}{\mJy}\right)^{2/5}\left(\frac{\nu}{10\GHz}\right)^{-1},\cr
B=&0.48 G\left(\frac{D(p)}{D(3)}\right)^{2/5}\cr
&\times\left(\frac{\sigSIC F_t}{10^{-12}\erg\cm^{-2}\sec^{-1}}\right)^{2/5}\left(\frac{f_{\nu,\rm abs}}{\mJy}\right)^{-2/5}\frac{\nu}{10\GHz}.
\end{align*}

We thank Eli Waxman, Alicia Soderberg and Ehud Nakar for useful suggestions. This work was supported by NASA through Einstein Postdoctoral Fellowship awarded by the Chandra X-ray Center, which is operated by the Smithsonian Astrophysical Observatory for NASA under contract NAS8-03060.


\begin{thebibliography}{01}
\bibitem[Scott 
\& Readhead(1977)]{Scott77} Scott, M.~A., \& Readhead, A.~C.~S.\ 1977, \mnras, 180, 539 
\bibitem[Slysh(1990)]{Slysh90} Slysh, V.~I.\ 1990, Soviet 
Astronomy Letters, 16, 339 
\bibitem[Chevalier(1998)]{Chevalier98} Chevalier, R.~A.\ 1998, 
\apj, 499, 810 
\bibitem[Chevalier 
\& Fransson(2006)]{Chevalier06} Chevalier, R.~A., \& Fransson, C.\ 2006, \apj, 651, 381 
%\bibitem[Svensson(1984)]{Svensson84} Svensson, R.\ 1984, MNRAS,
%209, 175
\bibitem[Rybicki \& Lightman(1979)]{Rybicki79} Rybicki, G.B. \& Lightman, A.P., Radiative Processes in Astrophysics, Jhon
WIley \& Sons, Inc.

*Bahcal Fellow, Einstein Fellow


\end{thebibliography}
\end{document}